\documentstyle[aps,prl,multicol,epsfig]{revtex} 
 
\begin{document} 
 
\tightenlines 
\title{Quantum computation using electrons trapped by surface acoustic waves} 
 
\author{ C. H. W. Barnes, J. M. Shilton and A. M. Robinson } 
\address{ Cavendish Laboratory, University of Cambridge, Madingley Road, 
Cambridge CB3 0HE, United Kingdom } 
 
\date{\today} 
\maketitle 
 
\begin{abstract} 
We describe in detail a set of ideas for implementing qubits, 
quantum gates and quantum gate networks in a semiconductor 
heterostructure device. Our proposal is based on an extension of 
the technology used for surface acoustic wave (SAW) based 
single-electron transport devices. These devices allow single 
electrons to be captured from a two-dimensional electron gas in 
the potential minima of a SAW. We discuss how this technology can 
be adapted to allow the capture of electrons in pure spin states 
and how both single and two-qubit gates can be constructed using 
magnetic and non-magnetic gate technology.  We give designs for 
readout gates to allow the spin state of the electrons to be 
measured and discuss how combinations of gates can be connected to 
make multi-qubit networks. Finally we consider decoherence and 
other sources of error, and how they can be minimized for our 
design. 
\end{abstract}

 
\begin{multicols}{2} 
\narrowtext 
 
\section{Introduction} 
 
The quantum mechanics of interacting many-particle systems 
presents an intractable problem for classical computers. The vast 
Hilbert space and the correlation between particles prohibits the 
exact simulation of all but the smallest systems by classical 
means. Thus, by manipulating and observing even a relatively small 
many-particle system a ``computation" could be performed which no 
classical computer would be capable of. Such a system, together 
with its measuring apparatus is referred to as a ``quantum 
computer" \cite{preskill,popescu}. Extensive research in this 
field has shown that quantum computers can be used to solve any 
computational problem, that is, they are capable of universal 
computation. However, it is clear that they are efficient at 
simulating quantum systems,\cite{Feynmann,Lloyd1,Zalka,Somaroo1}, 
and fast algorithms for factoring large numbers \cite{Shor} and 
searching databases \cite{Grover1,Grover2,Grover3} have been 
found. 
 
Two factors considerably simplify the design of a suitable 
many-particle system for universal quantum computation: it is 
sufficient to construct only a single line of two-state systems 
(so called qubits) which are weakly coupled 
\cite{Lloyd2,Divincenzo1,Sleator}; and it is not necessary to 
maintain global phase coherence throughout a computation 
\cite{Shor2,Steane,Chuang6,Chuang3,Chuang2,Cory2}. These 
simplifications are probably outweighed by the necessity that each 
qubit must be individually addressable, so that its pseudospin can 
be independently rotated through an arbitrary angle about any two 
chosen ``axes'', and by the requirement that the coupling between 
each pair of qubits must be accurately controllable. Useful 
qubits, together with suitable coupling mechanisms, can be found 
in almost every area of physics and chemistry and there are a 
growing number of attempts to demonstrate control over the 
rotation of and coupling between qubits 
\cite{Milburn,Barenco,Cirac,Pellizari,Turchette,Monroe,Gershenfeld,Chuang5,Havel,Bonadeo,Platzman,Kane1,Vrijen,Shnirman1,Averin,Ioffe,Makhlin,Nakamura1}. 
 
The proposal for quantum computation that we outline in this paper 
is in the semiconductor quantum dot category 
\cite{Burkard,Loss1,Loss2}. It is based on recent experiments in 
which a surface acoustic wave (SAW) is passed across a 
semiconductor heterostructure containing a two-dimensional 
electron gas (2DEG) and is incident on a depleted 
quasi-one-dimensional channel (Q1DC). In this experiment it is 
possible to capture a single electron from the 2DEG in each 
minimum of the SAW wave and transport them through the Q1DC 
\cite{Shilton96,Talyanskii97,Talyanskii98}. Our proposal is based 
on the assumption that by placing a series of $N$ Q1DCs in 
parallel, a SAW can be made to capture $N$ electrons in each of 
its minima, with one electron in each Q1DC. A single quantum 
computation would be performed by the $N$ electrons in a single 
SAW minimum as they are dragged through a pattern of magnetic and 
non-magnetic surface gates, which carry out single and two-qubit 
operations. At the end of the computation the electrons would then 
be channeled into a set of $N$ readout gates that would determine 
the final spin state of each electron. Each SAW minimum would 
contain an identical number of electrons prepared in an identical 
manner and would therefore perform the same computation. This 
would allow the readout gates to produce a measurable current 
representing the output of the quantum computer. Since the qubits 
would be carried along with the SAW, this proposal is an example 
of a ``flying qubit'' design \cite{Milburn,Ionicioiu}. We discuss 
these ideas based on using the electron spin to encode the qubits, 
but the idea could easily be extended to use the two states of a 
double quantum well, or the transverse states in laterally 
patterned parallel Q1DCs. 
 
The paper is organised as follows.  In section II we describe the 
SAW-based single-electron transport experiment, and discuss how 
SAWs incident on a Q1DC capture electrons.  We then discuss how a 
magnetic field would enable the electrons to be captured as qubits 
in prepared states. In section III we show how magnetic gating can 
be used for single qubit operations, and how non-magnetic gating 
can produce two-qubit operations. A simple model is used to detail 
the operation of the two-qubit gate. Section IV discusses the 
integration of the single and two-qubit gates to produce a C-NOT 
gate and larger networks of gates. Section V addresses the 
question of how the final state of the qubits can be read out and 
in section VI we consider possible sources of error and 
decoherence. 
 
\section{Production of single and multiple qubits} 
 
\subsection{Production of a quantised current using SAWs} 
 
Figure~\ref{ExptDevice} shows a schematic diagram of a SAW-based 
quantised-current device 
\cite{Shilton96,Talyanskii97,Talyanskii98}. It consists of a 
GaAs/AlGaAs heterostructure with NiCr/Al interdigitated 
transducers patterned either side of a central etch-defined mesa 
which contains a 2DEG. When a high frequency a.c. signal is 
applied to one of these transducers it produces a SAW, via the 
piezoelectric effect, which can be detected with the other 
transducer. As a SAW propagates across the mesa region, the 
traveling periodic electrostatic potential it produces drags 
electrons from the 2DEG along with it. For a typical SAW frequency 
of 3 GHz and an applied power of 10 dBm this produces a measurable 
current in the nanoAmp range. For the quantised current devices, 
the mesa is patterned so that it is split into two 2DEG regions 
(source and drain) connected by a narrow depleted Q1DC. The Q1DC 
can be formed either by surface Schottky gates, or by an etching 
technique developed by Kristensen {\it et al.} 
\cite{Kristensen1,Kristensen2} as shown in Fig~1. It has been 
shown experimentally \cite{Shilton96,Talyanskii97,Talyanskii98} 
that over certain ranges of SAW power and gate voltage, the 
current which is produced by the SAW becomes quantised in units of 
$ef$, where $e$ is the elementary electronic charge and $f$ is the 
SAW frequency. The lowest quantised value observed represents the 
transport of a single electron from the source to the drain 
reservoir in each SAW minimum through the Q1DC. The electrons 
travel through the Q1DC with a mean speed equal to the SAW 
velocity, which is 2700ms$^{-1}$ on the (001) plane of GaAs. 
Recent experimental results have demonstrated this effect with an 
accuracy of five parts in $10^4$ \cite{Cunningham}. 
 
\subsection{Discussion of capture process} 
 
At present, there is only an incomplete understanding of the 
mechanism by which electrons are captured and transported by the 
SAW through the Q1DC 
\cite{Aizin,Maksym,Flensberg,Pustilnik,Robinson}. This capture 
process is intrinsically time dependent; it involves the Coulomb 
interaction between many particles; and includes quantum 
mechanical confinement, tunneling and decoherence. 
 
Figures.\ref{CaptureB0}a,b show a contour plot of the effective 
potential in the Q1DC connecting the source and drain 2DEGs, and a 
cross-section through the same potential.  Both figures are 
schematic and shaded regions indicate electron occupation. These 
figures illustrate that as a SAW minimum approaches the entrance 
to the Q1DC, the SAW potential trough narrows and becomes a 
quantum dot. The size of this quantum dot implies that, at the 
point that it forms, it will be occupied by many electrons.  As it 
shrinks in passing through the Q1DC, electrons will be forced out. 
Electrons that are left in the dot are taken far from equilibrium 
with the source and drain reservoirs as they pass across the Q1DC. 
The distances and barrier heights between adjacent quantum dots 
are sufficiently large that the dots in the Q1DC can be considered 
to be independent as far as quantum-mechanical tunneling is 
concerned. 
 
From a classical perspective, the electrons that are most likely 
to be captured and transported through the Q1DC by the SAW will be 
those that are least energetic in the rest frame of the SAW. 
Quantum mechanically, the SAW minima will contain electrons in 
localised states and the SAW would not be expected to eject these 
electrons in favour of capturing those in higher energy states. 
This is borne out by classical simulations and by evaluations of 
the 1D single-particle time-dependent Schr\"odinger equation 
\cite{Robinson}. Figure~\ref{Wavefns} illustrates the 
quantum-mechanical capture process by showing the evolution of the 
wave functions of the first three bound states of a SAW minimum as 
it approaches and then passes through a Q1DC. The potential in 
Fig.~\ref{Wavefns} was chosen so that the mean current would 
correspond to the transport of 1.5 electrons per cycle. It can be 
seen in the upper sequence that the lowest state is simply carried 
along with the traveling SAW, even through the Q1DC. However, the 
second state is partially reflected as it arrives at the Q1DC and 
the third state is completely rejected. All higher energy states, 
including those that are delocalised in the bulk 2DEG region, are 
also reflected. 
 
The effect of the Coulomb interaction is important in the capture 
process because as quantum dots form and are brought to the Fermi 
energy, the screening between their trapped electrons is reduced. 
Once above the Fermi energy this screening appears only as an 
image charge in the 2DEG. The Coulomb interaction will therefore 
control the form of the time-dependent wave functions, the state 
energies, and ultimately the number of electrons left in a dot as 
it passes through the Q1DC. 
 
Quantum-mechanical decoherence is relevant to the capture and 
escape processes because it determines whether there is an integer 
number of electrons trapped in the dot or not, and it affects the 
quantum states of these electrons. If at some instant the energy 
of one of the electrons in a particular dot is too large for it to 
be contained by that dot - or large enough that tunneling becomes 
important - then the wave function of this electron will start to 
leak out of that dot. An electron which is in the process of 
leaving will initially propagate over the adjacent 2DEG 
maintaining coherence with the electrons left in the dot and will 
then lose this coherence as it interacts with conduction 
electrons. The decoherence will determine whether the electron 
leaves and will cause the dot to be left in a mixed state. As more 
and more electrons escape it is probable that the density matrix 
of the dot will become increasingly diagonal. For the case where 
finally only a single electron remains in the dot this would mean 
that its density matrix would simply be a half and half mixture of 
spin up and spin down. Such a mixed state is not suitable for 
quantum computation. 
 
\subsection{Pure states} 
 
For quantum computation it is necessary to provide electrons in 
pure states and this can be achieved by applying an external 
magnetic field to our device.  The simple application of an 
external magnetic field to an $N$ particle mixed state does not 
cause it to become a pure state unless it has time to relax by 
emitting phonons or photons: the field merely precesses the spin 
of each electron. However, it has a significant effect on 
which electrons are captured from the 2DEG by the SAW \cite{Robinson}. 
 The application of an external magnetic field polarises 
the electrons in the source 2DEG so that the low energy electrons 
all have the same spin.  The Lande g-factor in GaAs is 0.44 and 
the amplitude of the SAW electrostatic potential is estimated from 
experiments to be approximately 1meV 
\cite{Shilton96,Talyanskii97,Talyanskii98}. Therefore, the 
application of a magnetic field with strength $\sim$1T would 
ensure that all electrons captured would have the same spin. This 
is illustrated schematically in Fig.~\ref{CaptureB}. Once the 
split-gate voltage is increased to the point where only one 
electron is captured per cycle these electrons will be in pure 
spin states even though their orbital motion may be mixed. These 
electrons can therefore be used as spin qubits. 
 
In principle a large number of such qubits may be prepared 
simultaneously using this method by patterning large numbers of 
parallel Q1DCs with surface gates~\cite{Haubrich} or by etching. 
 
\section{Qubit gate operations} 
 
A quantum circuit consists of devices performing two kinds of 
operation on qubits. The first, the single qubit operation, acts 
to rotate the spin of a single qubit about an arbitrary direction 
in space. The second, the two qubit operation, entangles the spin 
wave functions of two adjacent qubits by altering the exchange 
coupling between them. A gate which produces such entanglement and 
is easy to produce within our scheme is the ``square root of 
swap'' gate \cite{Burkard,Loss1,Loss2}. 
 
\subsection{Single qubit gate} 
 
We can perform a single qubit operation on a SAW spin qubit by 
having it pass through the magnetic field from a local static 
magnet. During the time a qubit is in the field, its wave function 
will evolve according to the Zeeman term in Schr\"odinger's 
equation 
\begin{equation} 
\left[\matrix{S_z&S_x-iS_y\cr 
           S_x+iS_y&-S_z}\right] 
\left[ \matrix{\alpha \cr 
         \beta }\right] 
 = i { \partial \over \partial t} 
\left[ \matrix{\alpha \cr 
         \beta }\right] 
\end{equation} 
where $S_{x,y,z} = g\mu_BB_{x,y,z}/2$ and the qubit wave function 
has the form; 
\begin{eqnarray} 
|\psi \rangle &=& \alpha|\uparrow \rangle +\beta|\downarrow 
\rangle 
\end{eqnarray} 
For a constant external magnetic field this evolution is a 
precession of the qubit spin about the direction of the field. 
 
In principle, a local static magnet with a magnetic field that 
points in an arbitrary direction can be provided by a magnetic 
force microscope or by controlled ion beam deposition of a 
ferromagnetic material. However, for universal computation it is 
only necessary to make single qubit rotations about two 
independent axes and therefore it is sufficient to provide local 
magnetic fields that point in two orthogonal directions. As we 
have mentioned above, in order to provide pure states for the 
input, the device must be placed in an external field. This field 
may be used for one of the directions, and could be screened in 
regions where it is not needed by a superconducting material such 
as niobium.  The magnetic field for the other direction can be 
provided by a local magnetic split gate such as that shown in 
Fig.~\ref{OneQubit}.  If this is fabricated as a split ring, then 
stray fields will be reduced.  A material such as permalloy could 
be used since it has a large surface magnetisation ($\sim$1T) and 
exhibits shape anisotropy \cite{Gardelis} which can be exploited 
in setting the direction of the magnetisation. If the application 
of a global external magnetic field proved to be inconvenient for 
a particular circuit design, the input magnetic field and both 
directions of magnetic field for single qubit operations could be 
provided by magnetic surface gates. 
 
Figures \ref{Rab}a,b show how a qubit spin will precess as it 
passes through a magnetic split gate. They are calculated from the 
Zeeman contribution to the time-dependent Schr\"odinger equation 
eqn~1. For these figures, a constant field $B_z$ has been applied 
and a local field $B_y$ is switched on for a short time, 
representing the passage of a qubit through a magnetic split gate. 
In the regions where $B_y=0$ the external field $B_z$ causes the 
spin to precess around the $z$ axis without altering its component 
in this direction, so that $|\alpha/\beta|^2 = const$.  In the 
region where $B_y\neq 0$ this ratio oscillates at the Rabi 
frequency. These figures show that it is possible to shift the 
phase of the qubit components independently by altering the length 
of channel exposed to the external constant field, and to vary the 
relative amplitudes of the qubit components by varying the length 
of the magnetic gate.  A surface gate of length $\sim$220nm with a 
surface field of $\sim1$T will produce a $\pi/2$ rotation on a 
passing qubit spin.

\subsection{Two qubit gate} 
 
A two qubit operation can be achieved with surface patterning 
which causes two adjacent qubits, captured in the same SAW 
minimum, to be forced into tunnel contact.  Suitable gate 
patterning is shown schematically in Fig.~\ref{TwoQubit}. It 
consists of etched trenches defining an `X shape' with a central 
metallic surface gate which splits this shape into two channels 
when a negative bias is applied to it. The tunneling coefficient 
between adjacent dots is fully controllable using the bias on the 
central surface gate. For a low bias, the tunneling coefficient is 
essentially unity, and at a large negative bias it will be zero. 
For a finite transmission coefficient the wave functions of the 
two electrons in the two dots entangle via tunneling and the 
Coulomb interaction and then decouple leaving just one electron in 
each quantum dot. This represents a two-particle unitary 
transformation on the initial qubit states of the trapped 
electrons.

\subsection{A simple model} 
 
A number of models for a two-qubit gate made from two coupled 
quantum dots are given in~\cite{Burkard}. These models also apply 
to the two-qubit gate described above. Here, we discuss the basic 
operation of this gate in terms of the Hubbard model.  This model 
incorporates the possibility for double occupation of a single dot 
which we discuss later in the section on readout gates. The 
electrons in the channels shown in Fig.~\ref{TwoQubit} are 
sufficiently separated that they cannot tunnel between wells in 
the direction of propagation of the SAW. Away from the central 
tunnelling region, the wave functions in the upper and lower 
channels will be spin qubits of the form: 
\begin{eqnarray} 
|\psi_u \rangle &=& \alpha|\uparrow \rangle +\beta|\downarrow 
\rangle\\ 
 |\psi_l \rangle &=& \gamma|\uparrow \rangle 
+\delta|\downarrow \rangle 
\end{eqnarray} 
where $|\alpha|^2+|\beta|^2=1$ and $|\gamma|^2+|\delta|^2=1$.  The 
indices $u,l$ refer to the upper and lower channels respectively. 
The corresponding two-particle wave function will have the form; 
\begin{eqnarray} 
|\psi\rangle = \beta\delta|\chi_1\rangle + 
\alpha\gamma|\chi_2\rangle + 
 \alpha\delta|\chi_3\rangle + 
\beta\gamma|\chi_4\rangle 
\end{eqnarray} 
where 
\begin{eqnarray} 
|\chi_1 \rangle &=& 
c^\dag_{u\downarrow}c^\dag_{l\downarrow}|0\rangle \\ |\chi_2 
\rangle &=& c^\dag_{u\uparrow}c^\dag_{l\uparrow}|0\rangle 
\\ |\chi_3 \rangle &=& 
c^\dag_{u\uparrow}c^\dag_{l\downarrow}|0\rangle \\ |\chi_4 \rangle 
&=& c^\dag_{u\downarrow}c^\dag_{l\uparrow}|0\rangle 
\end{eqnarray} 
The operator $c^\dag_{i\sigma}$ creates an electron in dot $i$ 
with spin $\sigma$ from the empty state $|0\rangle$.  In the 
region where the two qubits entangle, we consider two extra states 
which represent the possibility for double occupation of one of 
the dots. 
\begin{eqnarray} 
|\chi_5 \rangle &=& 
c^\dag_{u\downarrow}c^\dag_{u\downarrow}|0\rangle \\ 
 |\chi_6 
\rangle &=& c^\dag_{l\uparrow}c^\dag_{l\uparrow}|0\rangle 
\end{eqnarray} 
Assuming that the electrons on each dot are tightly bound so that 
weak tunneling is permitted between adjacent dots in the same SAW 
minimum, and assuming an on site Coulomb interaction, the time 
dependent Schr\"odinger equation takes the form: 
\begin{equation} 
\left[\matrix{-2S&\quad&\quad&\quad&\quad&\quad\cr 
           \quad&2S&\quad&\quad&\quad&\quad\cr 
           \quad&\quad&0&\quad&\nu&\nu\cr 
           \quad&\quad&\quad&0&-\nu&-\nu\cr 
           \quad&\quad&\nu&-\nu&U&\quad\cr 
           \quad&\quad&\nu&-\nu&\quad&U}\right] 
\left[ \matrix{\chi_1 \cr 
         \chi_2 \cr 
         \chi_3 \cr 
         \chi_4 \cr 
         \chi_5 \cr 
         \chi_6 }\right] 
 = i { \partial \over \partial t} 
\left[\matrix{\chi_1 \cr 
         \chi_2 \cr 
         \chi_3 \cr 
         \chi_4 \cr 
         \chi_5 \cr 
         \chi_6 }\right] 
\end{equation} 
where $U$ is the on site Coulomb energy, $S$ is the Zeeman energy 
and $\nu$ is the single-particle tunneling energy between the two 
dots. The form of this Hamiltonian indicates that the states 
$\chi_1$ and $\chi_2$ evolve independently and are unaffected by 
the narrow tunnel barrier in the central region of the gate.  This 
derives from the fact that two-particle tunneling between dots 
necessarily involves double occupation.  Since the two electrons 
in these states have the same spin double occupation is forbidden 
by the Pauli exclusion principle and therefore no tunneling can 
occur. This assumes that there are no other single-particle states 
which can be occupied. The remaining part of the Hamiltonian has 
four eigenvalues of the form: 
\begin{eqnarray} 
\epsilon_1&=&-J={1 \over 2}\left( U - \sqrt { U^2 +16\nu^2} 
\right)\\ \epsilon_2&=&0 \\ \epsilon_3&=&U \\ \epsilon_4&=&U+J={1 
\over 2}\left( U + \sqrt { U^2 +16\nu^2} \right) 
\end{eqnarray} 
For $U$ and $\nu$ constant this implies that the dynamics of the 
system consist of a `slow' oscillation at the exchange frequency 
$J$ between states $\chi_3$ and $\chi_4$, via the double 
occupation states $\chi_5$ and $\chi_6$, and a `fast' oscillation 
of states $\chi_5$ and $\chi_6$ of frequency $U$. Fig.~\ref{HubbL} 
shows a characteristic time dependence of the states $\chi_3 ... 
\chi_6$. The upper and lower traces in black show the chosen time 
dependence of $U$ and $\nu$. The lines in grey show the occupation 
probabilities of the states $\chi_{3,..,6}$. Figure~\ref{RootSWP} 
shows the operation of the root of swap gate which may be 
constructed from the two-qubit gate Fig.~\ref{TwoQubit}. Within 
our notation, the swap operation is a unitary operation which 
leaves the probability amplitudes for the first two states 
unchanged $\chi_{1,2} \rightarrow \chi_{1,2}$ and swaps the 
amplitudes of the second two states $\chi_{3,4} \rightarrow 
\chi_{4,3}$. Root of swap is the square root of this unitary 
transformation.  For this figure, the Coulomb energy $U$ has been 
set to be much larger than the exchange energy $J$ and this has 
resulted in low occupation probabilities for the double occupation 
states $\chi_{5,6}$.  This is desirable for quantum computation 
since these states are not qubit states and their occupation 
results in a form of decoherence since only the qubit states are 
measured at the end of the computation. For $J=0.01meV$ the time 
for a swap operation will be approximately $\Delta t=0.2$ns. 
Since the SAW travels at 2700ms$^{-1}$ this would correspond to a 
gate length $L=$550nm. The exchange energy $J$ depends on the 
tunneling probability across the central barrier region and it can 
therefore be set to have any value. However, using such a low 
value would need to be weighed against the possibility for a 
potential modulation of the order $J$ in the central gate region 
and the necessity for the temperature to be sufficiently low.  A 
practical design would have as large a value for $J$ as possible, 
the restrictions being the smallest length of gate which could be 
made and the need to maximize $U$ to prevent double occupation. 
 
\section{Quantum gate networks} 
 
\subsection{The C-NOT gate} 
 
The single and two-qubit gates we have described constitute the 
only two gates necessary for the construction of universal logic 
in a quantum computer \cite{Lloyd2,Divincenzo1,Sleator}. These two 
gates can be used to construct the quantum C-NOT operation 
\cite{Burkard}, a useful building block in quantum computer 
designs (e.g. see ref~\cite{Barenco}). In terms of our gate 
patterning, the C-NOT operation has the form; 
\begin{eqnarray} 
U_{C-NOT} &=& \exp(i\pi S_2^y)\exp(-i\pi S_1^z/2)\exp(-i\pi 
S_2^z/2)\nonumber 
\\&&\quad U_{SW}^{1/2}\exp(i\pi S_1^z)U_{SW}^{1/2}\exp(-i\pi 
S_2^y/2) 
\end{eqnarray} 
The gate patterning necessary to produce this is shown in 
Fig.~\ref{CNOT}. The `horse-shoe' gates in the figure are used to 
indicate the necessity for screening stray fields from magnetic 
gates.  This may can also be achieved by using a superconducting 
material such as niobium. 
 
\subsection{Networks}

A schematic diagram of a quantum gate network is shown in 
Fig.~\ref{Network}. It consists of a large number of parallel 
Q1DCs. Each Q1DC has single-qubit gates placed along its length 
and pairs of Q1DCs are coupled by two-qubit gates. Each Q1DC 
terminates with a readout gate. A single computation is performed 
by each SAW minimum as it passes across the network dragging a set 
of $N$ qubits along with it. The qubits in each SAW minimum are 
prepared in the same way and perform the same computation.  At the 
readout stage, each qubit passes into a separate readout gate 
which has two outputs arranged such that the ratio of the currents 
passing into them reflects the relative spin along a particular 
direction. The SAW performs $\sim 3.0\times 10^9$ computations per 
second and each calculation contributes to produce a measurable 
output current. 
 
\section{Readout} 
 
Many proposals for quantum computers rely on single electron 
transistors for reading the state of the output 
qubits~\cite{Shnirman2,Kane1,Ionicioiu,Makhlin}. Such techniques 
are also possible here and would be necessary if for a particular 
algorithm the output from individual computations was needed. With 
our device design though we can also construct readout gates which 
measure the ratio $\langle |\alpha|^2 \rangle : \langle |\beta|^2 
\rangle$ for the qubits passing out of each Q1DC.  The brackets 
indicate an average of the qubit spin components over $\sim 
3\times 10^9$ quantum computations per second. Since each 
computation from each SAW minimum is nominally identical in our 
scheme, this ratio should be the same as that for the individual 
qubits exiting from each Q1DC. Such averaging may even compensate 
for decoherence and errors from random events in a given network. 
We present three ways for measuring these averages here. 
 
\subsection{Magnetic readout} 
 
The simplest readout idea conceptually is to use the spin-valve 
effect to measure the orientation of each output qubit using 
ferromagnetic ohmic contacts \cite{Gardelis}. Figure~\ref{Read1} 
shows a specific design. The two ohmic contacts are ferromagnetic 
and have magnetisations pointing in opposite directions so that 
they have different resistances to the two spin components of 
incoming qubits.  If the total width of these ohmic contacts is 
made less that the spin coherence length this should result in the 
currents which flow through them to ground reflecting the spin 
orientation of the output qubits. Such a readout gate could be 
calibrated using a set of qubits of known orientation.  The stray 
fields which the magnetic contacts produce may be compensated for 
with single-qubit gates prior to detection. 
 
\subsection{Double occupation readout} 
 
Fig.~\ref{Read2} shows a readout gate which is an adaptation of 
the two-qubit gate.  It operates by comparing a test qubit (upper 
channel) with an unknown qubit (lower channel). In the central 
region there is a gold surface gate to produce an adjustable 
tunnel barrier between the upper and lower channels. This enables 
entanglement of incoming qubit wave functions. If the test qubit 
has the same spin orientation as the unknown qubit then no 
tunneling will occur between the two channels in the central 
region and the two incoming electrons will simply pass into ohmic 
contacts A and B. However, if they have opposite spin, 
 tunneling can occur and there 
is a finite probability that the exiting dots will be doubly 
occupied (see fig~\ref{HubbL} ) either in the upper or lower 
channel. The gate is designed so that if this occurs one of these 
electrons can further tunnel across a narrow barrier to ohmic 
contact C. Thus the ratio of the currents passing into A,B and C 
will reflect the spin of the unknown electron relative to that of 
the known test electron.

\subsection{Stern-Gerlach readout gate} 
 
Figure~\ref{Read3} shows a readout gate which is based on the 
Stern-Gerlach effect.  The single magnetic gate produces a 
magnetic field which is spatially varying in the $y$-direction. 
Through the Zeeman term, this then produces an equal and opposite 
force $F_{SG}$ on the different spin components of the incoming 
qubits so that the ratio of the currents flowing into ohmic 
contatcs A and B should be representative of their spin components 
along the $y$-direction. Since electrons are charged particles, 
this readout gate would need to be designed to eliminate, or 
compensate for, the effect of the Lorentz force $F_{Lor}$. It is 
nominally significantly larger $F_{Lor}/F_{SG} = 
ev_{SAW}B/{1\over2}g\mu_B{\partial B_y/\partial y} \sim 10^8 
B/{\partial B_y/\partial y}$.  The Lorentz force deriving from the 
magnetic field in the $y$-direction acts in the direction 
perpendicular to the 2DEG plane, the $x$-direction, and is 
therefore compensated for by the 2DEG quantum well confinement. 
The Lorentz force deriving from any resultant component of the 
magnetic field in the $x$-direction will produce a force in the 
$y$-direction and will therefore alter the relative currents into 
A and B. This could be compensated for by introducing a second 
pole piece and applying a d.c. bias $V_{dc}$ between them to 
balance the force: $F=0=E+v_{SAW}B_{R}$, where $E=V_{dc}/s$ is the 
electric field between the pole pieces (separation $s$) and $B_R$ 
is the resultant magnetic field in the $x$-direction. For a 
separation between pole pieces of $s=$5$\mu$m a resultant field of 
$B_R=0.1$T could be compensated by a d.c. bias $V_{dc}=1.5$mV. The 
device would need to be calibrated by comparing the currents for 
polarized and unpolarized currents.

\section{Errors and decoherence} 
 
\subsection{Fabrication errors} 
 
The SAW quantum-gate network consists of a set of static metal or 
etched gates.  Errors in their patterning will lead to a 
quantum-gate network performing a different unitary transformation 
from the one intended. 
 
 Errors in the length of magnetic split gates will lead to either an over or an under rotation of 
spins for single qubit gates. In principle, this kind of error can 
be compensated for by splitting the magnetic gates into sections 
and applying different potential differences between pairs of 
split gates to move the electrons closer to or further from them 
and therefore modify the field that they see.  Each gate would 
need to be individually tuned in this way.  For a single gate it 
would be possible to design it to be sufficiently long so that 
such fabrication errors were irrelevant but for a network, such 
small errors would build up. 
 
For the two qubit gate the design is such that the length of the 
tunneling region is not crucial since, by applying a potential the 
the central metal gate, we have control over its height. 
 
Errors in the width of Q1DCs could lead to double occupation or 
occupation of complicated linear combinations of orbital states. 
Such errors can be compensated for by applying suitable biases to 
surface gates. 
 
If local patterned magnets are used then any design must take 
account of stray fields which they may introduce into other parts 
of the system.  Either these fields would need to be screened 
using superconducting surface gates or they would need to be 
incorporated into the circuit design. 
 
MBE-grown GaAs HEMT structures typically provide a clean system 
for the study of quantum effects.  However, some level of 
impurities and disorder is inevitable.  The presence of this 
disorder will have a similar effect to channel width variation and 
could be compensated for in the same way. However, additionally, 
if a trap state or other impurity is very close to a Q1DC it could 
even remove an electron or add an extra electron to a SAW minimum 
occasionally. Provided the error rate due to these mechanisms was 
low, it would be removed in the averaging of the readout gates. 
 
Johnson noise on surface gates will cause fluctuations in all 
gates. The degree of filtering applied to the voltage lines 
providing the potential to set these gates in an experimental 
system will have to be limited by the desired operation speed.  At 
3GHz gate potential fluctuations may not be a significant problem, 
and again such fluctuations will tend to be averaged out in the 
readout gates. 
 
Temperature effects may cause a severe problem.  The SAW 
single-electron devices require cooling to 1.2K for optimum 
operation, but the large powers applied to the transducers to 
produce the effect will inevitably cause local heating at one end 
of the device.  Whether this will affect device performance will 
have to be determined experimentally. 
 
Surface-gate patterning would need to be designed such that it did 
not significantly modify the SAW amplitude through screening. This 
can be compensated for by using a HEMT with a 2DEG which is 
sufficiently deep that surface screening is 
irrelevant~\cite{Aizin,Hoskins}. 
 
\subsection{Decoherence} 
 
Decoherence is not an intrinsic limiting factor to quantum 
computation since quantum error correction codes can be used to 
correct for its effects 
\cite{Shor2,Steane,Chuang6,Chuang3,Chuang2,Cory2}. For our design, 
implementing these codes would simply result in the necessity to 
use more C-NOT gates for the same computation. 
 
Semiconductor systems suffer from a large number of possible 
sources of decoherence. Any coupling between any line of $N$ 
qubits and any other system of particles could reduce its 
coherence.  The sources of decoherence for our system would 
include at least: coupling to other lines of $N$ qubits in 
neighbouring SAW minima, and to other electrons, in surface gates, 
on donor impurities and in the adjacent 2DEGs; piezoelectric 
coupling to phonon modes; coupling to nuclear spins; and coupling 
to radio-frequency photons. 
 
The problem with coupling to other lines of qubits could be 
minimised, or even eliminated, by using SAW transducer designs 
which produce strong SAW minima at multiples of the fundamental 
operation frequency, allowing the SAW waveform to be 
non-sinusoidal.  A pulsed mode operation is also possible, by 
switching the signal to the transducer at radio frequencies. These 
techniques could be used to separate adjacent SAW minima by 
distances larger than the SAW wavelength and therefore reduce 
crosstalk between them. 
 
The problem of decoherence of electrons in a quantum dot caused by 
interaction with nearby conduction electrons has been demonstrated 
in transport measurements \cite{Yacoby3,Yacoby2,Yacoby1,Buks}. 
These measurements indicate that even a weak capacitive coupling 
to a nearby system can reduce quantum coherence within a device. 
Avoiding such coupling is a matter of design, but utimately will 
be an intrinsic limitation to any semiconductor quantum gate 
design. 
 
Bulk spin resonance measurements~\cite{Kikkawa} show that the 
limiting factor for the spin lifetime in GaAs is phonon 
scattering.  However, even at 5K lifetimes of up to 100ns were 
found. In this time the SAW travels approximately 300$\mu$m, so 
given a typical gate size of a micron, it should be possible for a 
SAW minimum to traverse hundreds of single and two-qubit gates 
before the computation it is performing loses all coherence. 
 
Decoherence due to acoustic phonons in coupled quantum dots has 
been investigated experimentally \cite{Fujisawa}. They find that 
the tunneling process which occurs in these devices actually 
produces phonons, thereby reducing the coherence time.  Since the 
two-qubit gate in our scheme relies on tunneling between adjacent 
quantum dots, this may also prove to be a limitation on our 
design. However, since the majority of these phonons will be of 
very long wavelength, it should be possible to reduce their effect 
by placing the device in a suitable cavity \cite{Fujisawa}. 
 
The presence of radio frequency photons from the microwave signal 
used to generate the SAW can be significantly reduced by 
issolating the device with suitable screening. Since the radio 
frequency used in experiments with SAWs has a wavelength of 
approximately 10cm a cavity much smaller than this would work 
well. 
 
If the electron spins couple to the nuclear spins of the host 
crystal, then this provides an additional mechanism for 
decoherence. Such coupling has been observed in GaAs systems 
\cite{Berg,Kronmuller1,Kronmuller2} and its effect on double 
quantum dots has been described theoretically in \cite{Loss1}.

\section{Summary} 
 
We have suggested a way in which SAW devices can be used for 
quantum computation. The idea involves the capture of electrons in 
pure spin states from a 2DEG by a SAW to form qubits. Single-qubit 
operations are performed using local magnetic fields produced by 
magnetic surface gate patterning. Two-qubit operations are 
performed by allowing exchange coupling between qubits through a 
controllable tunnel barrier. A quantum computation with one of 
these devices would consist of a single SAW minimum dragging 
 a line of $N$ qubits through a patterned array of 
single and two qubit gates.  Each SAW minimum would perform the 
same computation. A number of different schemes for reading out 
the final state of qubits have been proposed: magnetic ohmic 
contacts; double occupation in two-qubit gates; and the 
Stern-Gerlach effect. The major problem problem for this kind of 
implementation will be decoherence arising from coupling with: 
other electrons; radio frequency photons; phonons; and nuclear 
spins. However, these problems are generic to any semiconductor 
quantum computation proposal and we believe that the repetition 
(three billion time per second) which our implementation allows 
could be exploited to counteract these problems. 
 
{\it Acknowledgements:} We are grateful to C.~J.~B.~Ford, D.~P.~ 
DiVincenzo, D.~Loss, G.~Milburn and G.~Burkard for instructive 
discussions and encouragement in this work.  CHWB acknowledges an 
Advanced Fellowship from the EPSRC.

\begin{figure} 
\begin{center} 
\caption{ Schematic diagram 
showing an experimental device which produces a quantised 
acoustoelectric current.} \label{ExptDevice} 
\end{center} 
\end{figure} 
 
\begin{figure} 
\begin{center} 
 \caption{ (a) Schematic 
diagram showing the effective potential due to a SAW passing 
across a Q1DC; (b) potential through the centre of (a), parallel 
to the Q1DC. Shaded regions in (a) and (b) indicate electron 
occupation.} \label{CaptureB0} 
\end{center} 
\end{figure} 
 
\begin{figure} 
\begin{center} 
 \caption{Time-dependent 
solutions to 1D single-particle Schr\"odinger equation for the 
first three states of a particular SAW minimum. The squared moduli 
of the wave functions as a function of $z$ are shown together with 
the effective potential. The time sequence is from left to right.} 
\label{Wavefns} 
\end{center} 
\end{figure} 
 
\begin{figure} 
\begin{center} 
 \caption{Schematic diagram 
showing the spin up and spin down occupations of the 1D SAW 
potential in a finite magnetic field. } \label{CaptureB} 
\end{center} 
\end{figure} 
 
\begin{figure} 
\begin{center} 
 \caption{Schematic diagram 
showing a magnetic split gate for performing a single qubit 
operation. Arrows indicate the direction of the magnetisation of 
the split gates.} \label{OneQubit} 
\end{center} 
\end{figure} 
 
\begin{figure} 
\begin{center} 
\caption{ (a) Magnitude of the 
two components of a SAW qubit as a function of time; (b) 
corresponding probability amplitudes as it passes through a 
magnetic split gate. The time dependences of the fields $B_z$ and 
$B_y$ are indicated on the right axis.} \label{Rab} 
\end{center} 
\end{figure} 
 
\begin{figure} 
\begin{center} 
\caption{ Schematic diagram 
showing a design for a two-qubit quantum gate.  Etched trenches 
define an `X shape' with controlling 2DEGs above and below and in 
the center there is a gold surface gate.} \label{TwoQubit} 
\end{center} 
\end{figure} 
 
\begin{figure} 
\begin{center} 
\caption{Absolute values of 
the probability amplitudes for states $\chi_{3,..,6}$ of the 
two-qubit gate in Fig.~\ref{TwoQubit} as a function of time. The 
values of $U$ and $\nu$, right axis, are chosen to show the 
generic behaviour of such a gate.} \label{HubbL} 
\end{center} 
\end{figure} 
 
\begin{figure} 
\begin{center} 
\caption{ Root of swap 
operation for two-qubit gate shown in Fig~7.} \label{RootSWP} 
\end{center} 
\end{figure} 
 
\begin{figure} 
\begin{center} 
\caption{ Design for a C-NOT 
gate based on equation~17. Grey lines indicate etched trenches 
defining two parallel Q1DCs. Horse shoe shapes represent magnetic 
split gates and black rectangles indicate `root of swap' 
operations.} \label{CNOT} 
\end{center} 
\end{figure} 
 
\begin{figure} 
\begin{center} 
\caption{ Schematic diagram 
showing the gate pattern layout for a SAW quantum gate network. 
Black and white vertical lines represent the SAW effective 
potential. The network of grey lines represents a set of Q1DC. 
Black dots represent qubits.  White squares on the right hand side 
represent readout gates.} \label{Network} 
\end{center} 
\end{figure} 
 
\begin{figure} 
\begin{center} 
\caption{Schematic diagram 
showing a readout gate which uses magnetic ohmic contacts.  The 
Q1DC is defined by two etched trenches and is controlled by 2DEGs 
above and below.  The magnetic ohmic contacts are placed in a 
widened part of the Q1DC.} \label{Read1} 
\end{center} 
\end{figure}

\begin{figure} 
\begin{center} 
\caption{ Schematic diagram 
showing a readout gate which exploits the possibility of double 
occupation in a two-qubit operation.} \label{Read2} 
\end{center} 
\end{figure} 
 
\begin{figure} 
\begin{center} 
\caption{ A schematic diagram 
showing a readout gate which uses the Stern-Gerlach effect.} 
\label{Read3} 
\end{center} 
\end{figure} 
 
\end{multicols} 
 
\end{document}